\documentclass[pra,twocolumn,showpacs,nofootinbib,aps,amssymb]{revtex4}
\usepackage{graphicx}
\usepackage{amsmath}
\usepackage{times}

\begin{document}
\title{Dynamics of a period-three pattern loaded Bose-Einstein condensate in
an optical lattice}
\author{A.-M. Rey$^{1,2}$, P. B. Blakie$^2$, and Charles W. Clark$^2$}
\affiliation{$^1$Institute for Physical Science and Technology,
University of Maryland, College Park, MD 20742}
\affiliation{$^2$Electron and Optical Physics Division,
National Institute of Standards and Technology,
Technology Administration,
U.S. Department of Commerce,
Gaithersburg, MD 20899-8410}
\date{\today}

\begin{abstract}
We discuss the dynamics of
a Bose-Einstein condensate initially loaded into every third site of an optical lattice
using a description based upon the discrete nonlinear Schr\"odinger equation.
An analytic solution is developed for the case of a periodic initial condition and is compared with numerical simulations for more general initial configurations.
 We show that  mean field effects in this system can
cause macroscopic quantum self-trapping, a phenomenon already
predicted for double well systems. In the presence of a uniform
external potential,  the  atoms exhibit generalized Bloch
oscillations which can be interpreted in terms of the interference of three different
Bloch states. We also discuss how the momentum distribution of the system can be
used as experimental signature of the macroscopic self trapping effect.
\end{abstract}
\maketitle

\section{Introduction}
An optical lattice is a periodic potential formed by the ac-starkshift at
the intersection region of two far detuned laser fields.
A Bose-Einstein condensate loaded into such an optical
lattice is a useful system for studying and controlling
the dynamics of ultra-cold atoms. In this system experimental studies in
the meanfield regime have provided elegant demonstrations of band structure
\cite{Anderson1998a,Greiner2001a,Denschlag2002a} and quantum chaos
\cite{Hensinger2001a}, and have considered the effects of
nonlinearity on the condensate dynamics
\cite{Morsch2001a,Burger2001a,Morsch2002a}.
By suitably loading a condensate into deep
optical lattices, number squeezing \cite{Orzel2001a} and
the Mott-Insulator transition \cite{Greiner2002a} have been observed.

In this paper, we investigate a Bose-Einstein condensate loaded
into every third site of an optical lattice, motivated by the recent
experimental realization of this system by the NIST group
\cite{Porto2002a}. In that experiment a combination of two
independently controlled lattices was used to prepare the
condensate into every third site of a single lattice. Briefly the
procedure consists of  loading a condensate into the ground band
of lattice with periodicity $3a$, so that the condensate is well
localized in the potential minima of this lattice.  A second
lattice of periodicity $a$ which is  parallel to the first
lattice, is then ramped up so that the superimposed light
potentials form a \emph{super-lattice} of period $3a$. For the ideal case
both lattice potentials are in-phase, and the addition of the second
lattice will not shift the locations of the potential minima from those of the
first lattice alone and the condensate will remain
localized at these positions. Finally, by removing the first
period-$3a$ lattice on a time scale long compared to band
excitations, but short compared to the characteristic time of
transport within the lattice, the condensate will be left in every
third site of the period-$a$ lattice.

A condensate loaded in this way is not an eigenstate
of the final period-$a$ lattice, and the condensate will
continue to evolve in the final system.
Outside the strongly correlated regime (where the depletion becomes
large \cite{ana2002a}) the Gross-Pitaevskii equation
is expected to give a good description of the condensate
dynamics, and has been applied
to the lattice system \cite{Choi1999a,Bronski2001a}.
In our system we assume the lattice is sufficiently
deep that a tight binding description is applicable,
and the Gross-Pitaevskii equation can be reduced to a
discrete nonlinear Schr\"odinger equation involving
only an onsite nonlinear interaction and nearest
neighbor tunneling. For the periodic initial condition
of equal wavefunction amplitude every third site and
zero at all others,  symmetry arguments can be used to
reduce the wavefunction evolution to a two mode problem
(analogous to a double well system with an energy offset)
for which an analytic solution of the dynamics can be given.
We show that for large ratios of the interatomic interaction
strength to tunneling energy the condensate evolves with
self-maintained population
imbalance, whereby the condensate population
tends to remain localized in the initially occupied lattice
sites.  A similar phenomenon has been studied in double
wells systems \cite{Smerzi1999a, Marino1999a,
Smerzi1999b,Smerzi2000a, Giovanazzi2000a}.
We show that the momentum distribution of an interacting
condensate changes in time in a manner which can be related
to the spatial tunneling of condensate, and would be a
suitable experimental observable.

Under the influence of an external force a Bloch state will
accelerate until it reaches the zone edge where it will
Bragg scatter to the opposite side of the Brillouin zone.
This periodic motion, known as Bloch oscillations, has
been observed in systems of cold atoms \cite{Dahan1996a} and
Bose-Einstein condensates in optical lattices
\cite{Morsch2001a,Denschlag2002a}.  To illustrate how a
linear potential affects the motion of a pattern loaded
condensate we find analytic solutions for the noninteracting
case with periodic initial condition and show how the dynamics
for this system can be interpreted in terms of the interference
of three Bloch states of the lowest band Bloch oscillating in unison.
 We present numerical results for more general (non-periodic) initial
 conditions and consider characteristic properties of the momentum
 distribution.

In Section \ref{secTB} we start by reviewing the basic formalism
of a Bose-Einstein condensate in an optical lattice in the tight
binding approximation. In
Section \ref{sectionnoextpot} we study the dynamics for
the case when only the
optical potential is present. We discuss a periodic
model and compare its predictions with numerical simulations for more general
initial conditions. In
section \ref{secextpot} we consider the dynamics of a Bose-Einstein
condensate in the lattice with
the presence of  constant external force.

\section{Tight binding description of a BEC in an optical lattice}
\label{secTB}
The dynamics of a condensate in the lattice can be modelled by the
1D Gross-Pitaevskii equation

\begin{eqnarray}
i\hbar \frac{\partial \Phi }{\partial t}&=&-\frac{\hbar ^{2}}{2M}\frac{%
\partial ^{2}\Phi }{\partial x^{2}}+  V_{0}\sin^2
\left(\frac{%
2\pi }{\lambda }x \right)\Phi \nonumber \\
&&+ V_{\rm{Ext}}(x) \Phi +N\frac{4\pi \hbar ^{2}a_{1}}{M}\left| \Phi \right|
^{2} \Phi ,\label{GPE}
\end{eqnarray}
where $V_{\rm{Ext}}(x)$ is the external potential, $V_0$ is the
depth of the optical potential which is determined by the
intensity of the laser beams, $\lambda $ is the wavelength of the
lasers, $a_{1}$ is the 1D renormalized scattering length with
dimension of an inverse length (which we obtain by requiring that
the value of the chemical potential for the 3D system matches the
1D  chemical potential \cite{Chiofalo2000a}, $M$ is the atomic
mass and $N$ the total number of atoms.

In the tight-binding approximation, valid if the height of the
lattice is higher than the chemical potential at each well, the 1D
condensate order parameter can be expanded in a Wannier basis
 keeping only the lowest band
\cite{Ziman1964}:

\begin{equation}
\Phi (x,t)=\sum_{n}\Psi _{n}(t)\phi _{n}(x),\label{dos}
\end{equation}
where $\phi _{n}(x)=\phi (x - na)$ is the condensate Wannier
function centered on the
n$^{th}$ lattice site, with $\int dx\,\phi _{n}(x)\phi _{n+1}^{\ast
}(x)$ $%
=0$ and $\int dx|\phi _{n}(x)|^{2}=1$, $a = \lambda/2$
is the lattice constant, and $\Psi
_{n}(t)$ is the n$^{th}$ amplitude. Replacing this ansatz in Eq.
\ref{GPE}, the Gross-Pitaevskii equation reduces to the discrete nonlinear Schr\"odinger
equation and the $\Psi _{n}$ satisfy:
\begin{equation}
i\hbar\frac{\partial \Psi _{n}}{\partial t}=-J(\Psi _{n-1}+\Psi
_{n+1})+(\epsilon_{n}+ N U |\Psi _{n}|^{2})\Psi _{n},
\label{SIdnlse}
\end{equation}
where $U =(4\pi \hbar ^{2}a_{1}/M)\int dx|\phi
_{n}|^{4}$ is the strength of the on-site repulsion of two atoms
occupying the lattice site $n$,
$%
\epsilon_{n}=\int V_{\rm{Ext}}(x)|\phi _{n}|^{2}dx$ describes the
energy offset of each lattice site $n$, and $J=\int \phi
_{n+1}^{\ast }[-(\hbar
^{2}/2M)(
\partial^2 /\partial x^2) +V_{0}\sin^2
(
2\pi x/\lambda)]\phi _{n}dx$ is the hopping matrix
element between adjacent sites. As shown in Ref.
\cite{Trombettoni2001a,Trombettoni2001b} , Eq. (\ref{SIdnlse}) can
be seen as an equation of motion ${i \hbar \partial \Psi _{n}}/{\partial
t}=%
{\partial H}/{\partial \Psi _{n}^{\ast }}$, derived from a
Hamiltonian function $H$ given by

\begin{equation}
H=\sum_{n}-J\left( \Psi _{n}^{\ast }\Psi _{n+1}+\Psi _{n+1}^{\ast
}\Psi _{n}\right) +\epsilon_{n}|\Psi _{n}|^{2}+\frac{NU}{2}|\Psi
_{n}|^{4},\label{HamilT}
\end{equation}
with $i\Psi _{n}^{\ast }$ and $\Psi _{n}$
treated as canonically conjugate
variables. Both the Hamiltonian $H$ and the norm
$\sum_{n}|\Psi_n|^2=1$ are conserved quantities.

The major interest of this paper is in the tunneling properties of
the condensate in the lattice, and in the noninteracting case the
time scale for tunneling is determined by the hopping matrix
element $J$. For this reason it is convenient to define a new
scale of time $\tau=Jt/\hbar$, and
energies $E_n=\epsilon_n/J$
and coupling constant $\Lambda= N U/J$. In terms of these new variables
the
tight binding evolution equation (\ref{SIdnlse}) takes the form
\begin{equation}
i\frac{\partial \Psi _{n}}{\partial \tau}=-(\Psi _{n-1}+\Psi
_{n+1})+(E_{n}+ \Lambda |\Psi _{n}|^{2})\Psi _{n}, \label{tres}
\end{equation}

\section{Case of no external potential}
\label{sectionnoextpot}
\subsection{The case of periodic initial
conditions: reduction to a two mode system}
The stationary states and other aspects of the dynamics
of Eq. (\ref{tres}) have been discussed by others
previously \cite{Trombettoni2001b,Trombettoni2001a,Wu2002a}.  We are motivated
by recent experiments to treat the
case of the particular nonstationary state described in the
introduction, in
which a condensate is initially
loaded into every third site of an optical
lattice.  We treat first a model case
in which no external potential is present,
($ E_{n}=0$), and in which the initial occupancies
of each third site are the same, and in which the
condensate initially has a uniform phase.  At $\tau=0$,
the amplitudes $\Psi _{n}(\tau)$ are given by
\begin{eqnarray}
\Psi_{3n}(0)&=&\sqrt{\rho}\label{IC1},\\
\Psi _{3n+1}(0)&=&\Psi_{3n+2}(0)=0,\label{IC2}
\end{eqnarray}
where $N\rho$ is the initial number of atoms per
occupied site.  This initial condition is homogeneous
in the sense that
each occupied site has the same amplitude and phase along the
length of the lattice. For an infinite lattice,
or one with periodic boundary conditions,
the amplitudes for all
initially occupied sites ($\Psi _{3n}$) evolve
identically in time, and
the amplitudes for the initially unoccupied sites
satisfy $\Psi_{3n+1}(\tau) = \Psi _{3n+2}(\tau)$
for all $\tau$ and all $n$. This allows us
to reduce the full set of
equations (\ref{tres}) to a set of two
coupled equations
\begin{eqnarray}
i\frac{\partial \Psi _{0}}{\partial \tau} &=&-2\Psi _{1}+\Lambda
|\Psi
_{0}|^{2}\Psi _{0},  \label{cinco} \\
i\frac{\partial \Psi _{1}}{\partial \tau} &=&-(\Psi _{1}+\Psi
_{0})+\Lambda |\Psi _{1}|^{2}\Psi _{1},  \label{seis}
\end{eqnarray}
where $\Psi _{3n} \equiv \Psi _{0}$ and
$\Psi _{3n+1} = \Psi _{3n+2} \equiv \Psi _{1}$ for all $n$.
The normalization condition is
\begin{equation}
|\Psi _{0}|^{2}+2|\Psi _{1}|^{2}=\rho.  \label{norm}
\end{equation}
The Hamiltonian function, $H$, of this system
is (from Eq. \ref{HamilT})
\begin{eqnarray}
H &=&\frac{J}{\rho}\Big( -2(\Psi _{0}^{\ast }\Psi _{1}+\Psi _{0}\Psi
_{1}^{\ast })-2|\Psi _{1}|^{2}+\frac{\Lambda }{2}|\Psi _{0}|^{4}
\nonumber\\
&& +\Lambda|\Psi _{1}|^{4}\Big)  \label{hamil} \\
&=& \frac{J \Lambda \rho}{2}.  \nonumber
\end{eqnarray}

  By writing $\Psi _{0}$=$\psi _{0}\sqrt{ \rho}$
and $\Psi_{1}$=$\psi _{1}\sqrt{\rho/2}$,
we can transform Eqs.
(\ref{cinco})-(\ref{seis}) to the form
\begin{eqnarray}
i\frac{\partial }{\partial \tau}\left(
\begin{array}{c}
\psi _{0} \\
\psi _{1}
\end{array}
\right)  &=&\left(
\begin{array}{cc}
\gamma|\psi _{0}|^{2} & -\sqrt{2} \\ -\sqrt{2} & \gamma\frac{|\psi
_{1}|^{2}}{\sqrt{2}}-1
\end{array}
\right) \left(
\begin{array}{c}
\psi _{0} \\
\psi _{1}
\end{array}
\right) ,\label{nl2state}
\end{eqnarray}
where $\gamma=\Lambda \rho$ is the ratio of on-site repulsion to
tunneling energies, and the normalization condition (\ref{norm})
is now $|\psi _{0}|^{2}+|\psi _{1}|^{2} =1$. The factor of
$\sqrt{2}$ difference in the definition of $\psi_0$ and $\psi_1$
arises because $\psi_1$ represents the amplitude of the two
initially unoccupied sites. With this factor incorporated,
the  matrix appearing in Eq. (\ref{nl2state})
is explicitly Hermitian. We note this
equation of motion is identical
to that for a condensate in a double well trap in the two mode
approximation \cite{Milburn1997a,Smerzi1997a}.

We note in passing that a similar reduction,
to a system of $[m/2] + 1$ equations,
exists for lattice systems that are
loaded such that only every $m$-th site is initially
occupied.

\subsection{Solution of the equations of motion}
It is convenient to write the lattice site amplitudes appearing in
Eqs. (\ref{cinco}) and (\ref{seis}) as $\Psi _{0}$=$ fe^{i\theta
_{0}}\sqrt{\rho}$ and $\Psi _{1}$= $ge^{i\theta _{1}}\sqrt{\rho}$,
where $f, g$, and $\theta$ are real.
By introducing the
phase difference $\phi =\theta _{0}-\theta _{1}$,
Eqs. (\ref{cinco}), (\ref{seis}) and (\ref{hamil}) can be recast
as
\begin{eqnarray}
\stackrel{.}{f} &=&2g\sin \phi ,  \label{eqpef} \\
\stackrel{.}{g} &=&-f\sin \phi ,  \label{eqpeg} \\
\frac{\gamma}{2} &=&-4fg\cos \phi +\frac{\gamma}{2}\left(
f^{4}+2g^{4}\right) -2g^{2},  \label{eqpeh}
\end{eqnarray}
The analytic solutions of Eqs. (\ref{eqpef})-(\ref{eqpeh}), found
using a procedure similar to that presented by Raghavan \emph{et
al.} \cite{Smerzi1999a}, can be expressed in terms of
Weierstrassian elliptic functions $\wp (\tau; g_{2},g_{3})$ \cite{AS64}.
The
analytic solutions are
\begin{eqnarray}
f(\tau) &=&\sqrt{1-\frac{24}{12\wp (\tau;g_{2},g_{3})+(9+2\gamma
+\gamma ^{2})},}
\label{solu} \\
g(\tau) &=&\sqrt{\frac{1-f(t)^{2}}{2},}\label{solug}
\end{eqnarray}
where the parameters g$_{2}$ and g$_{3}$ are given by
\begin{eqnarray}
g_{2}&=& \left(81-14\gamma ^{2}+4\gamma ^{3}+\gamma
^{4}\right)/12,\\
g_{3}&=&\left(729+243\gamma ^{2}-46\gamma ^{3}-15\gamma
^{4}+\gamma ^{6}\right)/{216}.
\end{eqnarray}
The solutions $f(\tau)$ and $g(\tau)$ are oscillatory functions
whose amplitudes and common period, $T(\gamma)$,
are determined by the parameter $\gamma$
(see Figs. 1 and 2). It is useful to qualitatively divide this
behavior into two regimes, separated by $\gamma=2$.  Analysis of
Eqs. (\ref{eqpef}) - (\ref{eqpeh}) shows that
$f(\tau_0)=g(\tau_0)$ for
some value of $\tau_0$ when $\gamma \le 2$, and
for $\gamma>2$, $f(\tau) > g(\tau)$
for all $\tau$.

\subsubsection{The tunneling dominated regime}
For $\gamma\lesssim2$, we find that the oscillation
period is essentially
constant (see Fig. 1).
\begin{figure}[!tbh]
{\centering \includegraphics[width=3.2in]{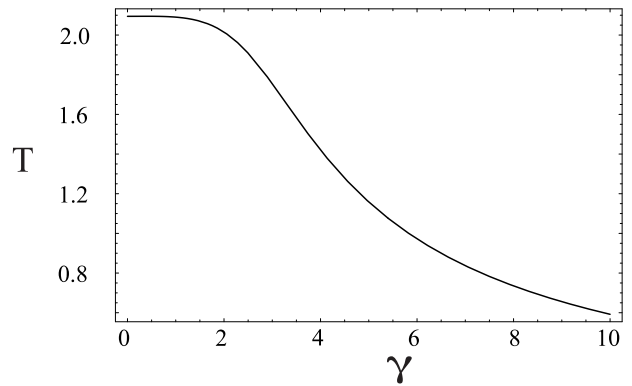}
\par}
\caption{Oscillation period (in units of $\hbar/J$)  as a
function of the interaction strength $\gamma$. \label{fig1}}
\end{figure}

In this case the role of interactions is relatively small, and the
behavior can be approximately understood by taking $\gamma=0$, in
which case the matrix of Eq. (\ref{nl2state}) is constant in time.
The equations of motion in this case are equivalent to those of a
two-state Rabi problem \cite{CohenTannoudji1977}, where the two
levels are coupled by a Rabi frequency of strength $\sqrt{2}$,
which is detuned from resonance by $-1$. This system will undergo
Rabi oscillations whereby atoms periodically tunnel from the
initially occupied site into the two neighboring sites. Because
the coupling is detuned from resonance the transfer of populations
between wells is incomplete, with $|\psi_1|^2$ attaining a maximum
value of $8/9$. The Rabi model predicts that the cycling frequency
between the levels is equal to the difference between the
eigenvalues of the matrix of matrix of Eq. (\ref{nl2state}), which
gives the period of oscillation as $T_{\rm{Rabi}}=2\pi/3$ in units of $\hbar/J$
(see Fig.  \ref{fig1}).

\subsubsection{Interaction dominated regime}
\begin{figure}[!tbh]
{\centering \includegraphics[width=3.2in]{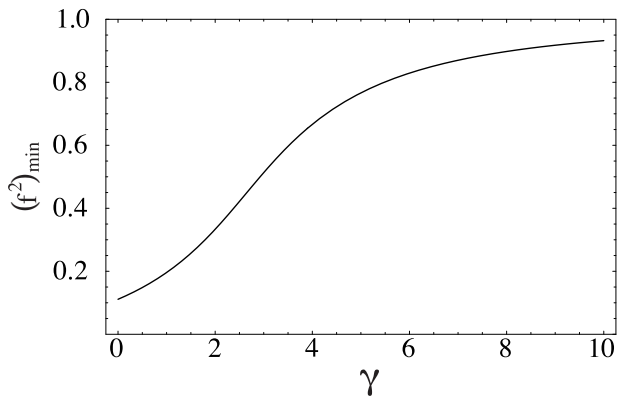}
\par}
\caption{Minimum value of $f^{2}$ during an oscillation period as a function of $\gamma$. As
$\gamma$ increases the population imbalance between wells increases (see text).  \label{fig2}}
\end{figure}

The effect of interactions on the mean-field dynamics is to cause
the energies of the initially occupied sites to shift relative to
those of the unoccupied sites. As $\gamma$ increases and this
energy shift increases relative to the strength of coupling
between sites, the tunneling between sites occurs at a higher
frequency, but with reduced amplitude. The population of the
initially occupied sites becomes  self trapped by the
purely repulsive pair interaction, which in the context of a
double well system has been called ``macroscopic quantum self
trapping'' \cite{Smerzi1997a}. This is demonstrated quantitatively
in Fig. \ref{fig2} where we plot the minimum value of $f^2$
occurring during the oscillation as a function of $\gamma$. In
contrast to the tunneling dominated regime, where tunneling
periodically populates all sites equally, the condensate tends to
be localized on the initially occupied sites in the
interaction-dominated regime.

\subsection{Momentum space dynamics}
Typically the spacing between individual wells in an optical
lattice is too small to resolve the localized density
distributions of atoms in neighboring sites using standard imaging
techniques.

The momentum distribution is a more convenient observable which
approximately corresponds to the expanded spatial distribution of
the released condensate. Here we calculate the momentum dynamics
of the condensate loaded into every third site of an optical
lattice, and show how this relates to evolution of the spatial
amplitudes given in Eqs. (\ref{solu}) and (\ref{solug}).

In the tight binding approximation the condensate order parameter
${\Phi}(x,\tau)$ is expressed as a sum over the lattice sites
(\ref{dos}). Because of the periodicity of the system, the
momentum space wavefunction, which we denote as
$\tilde{\Phi}(k,\tau)$, is expressible as the Fourier series
\begin{equation}
\tilde{\Phi}(k,\tau) =\sum_{m}\Psi _{m}(\tau)e^{-ikam}\chi (k),
\end{equation}
where \begin{equation} \chi (k)=\frac{1}{\sqrt{2\pi
}}\int_{-\infty }^{\infty }e^{-ikx}\phi _{0}(x)dx.
\end{equation}

To compute the momentum distribution,
we invoke the identity
$\sum_{n=0}^{N_{\rm s} -1}e^{ikna}=N_{\rm s }\delta
_{k, 2\pi m/a}$, where $N_{\rm s }$ is the number
of lattice sites,
and $m$ an integer. Since there are only
two independent amplitudes in the set $\{\Psi_m\}$, we find
that
\begin{eqnarray}
\tilde{\Phi}(k,\tau) &=&\sqrt{3/\rho}c_{m}(\tau)\chi _{m}\delta _{k,qm/3},
\nonumber\\
\chi_{m} &=&\chi(qm/3), \\
c_{m}(\tau) &=&\sqrt{\frac{1}{3 \rho}}\, \left( \Psi _{0}+\Psi
_{1}e^{-iqm/3}+\Psi _{2}e^{-i2qm/3}\right) , \nonumber
\end{eqnarray}

where $q$ is the reciprocal lattice vector $q=2\pi /a$. The
momentum distribution of the system has very sharp peaks of
relative amplitude $|c_m|^2$ at momentum $k=qm/3$, arising from
the $3$-lattice site spatial periodicity of the condensate
wavefunction. In addition, $\chi_m$ describes a slowly varying
envelope determined by the localization of the Wannier states at
each lattice site.

\begin{figure}[!tbh]
{\centering \includegraphics[width=3.2in]{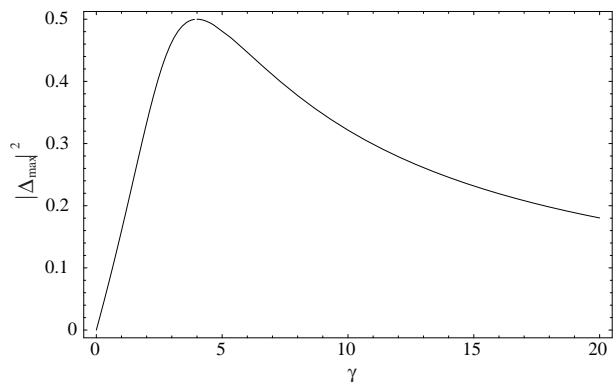}
\par}
\caption{Maximum contrast of the Fourier components as a function
of $\gamma$.  The maximum contrast is defined as  $\Delta _{\max
}\equiv(|c_{1}|^{2}-|c_{0}|^{2})_{\max}$ with the maximum value occurring  when $f^2$ is at its minima.  \label{fig3}}
\end{figure}

Using the analytic solutions for $f$ and $g$, and Eq.
(\ref{eqpeh}), we obtain only two independent Fourier amplitudes
\begin{eqnarray}
|c_{3n}(\tau)|^{2} &=&\frac{1}{3}\left( 1+\frac{\gamma
}{4}(3f^{2}+1)(f^{2}-1)%
\right) ,  \label{foam} \\
|c_{3n+1}(\tau)|^{2} &=&\frac{1}{3}\left( 1-\frac{\gamma }{8}%
(3f^{2}+1)(f^{2}-1)\right)  \label{foam2} \\
&=&|c_{3n+2}(\tau)|^{2}
\end{eqnarray}
In the reduced zone scheme, where we only consider momenta in the
range $k\in [-q/2,q/2]$,  the momentum wavefunction then consists
of three peaks corresponding to Bloch states of quasimomenta
$0,\pm q/3$. The identical behavior of $|c_{3n+1}|^2$ and
$|c_{3n+2}|^2$ means that the $\pm q/3$ peaks always have the same
intensity. If interactions between the atoms are ignored (i.e.
$\gamma=0$), the momentum components are constant in time (see
Eqs. (\ref{foam}) and (\ref{foam2})), even though tunneling occurs
between the lattice sites. However, when interactions are
considered, the momentum intensities explicitly depend on the
occupations of each site and will vary in time when tunneling
occurs. The magnitude of the time variation of the $|c_n|^2$ is
proportional to $\gamma$, but will reduce for sufficiently large
values of $\gamma$, where the self-trapping effect causes the
tunneling between lattice sites to stop (i.e. $f^2\approx1$ at all
times). In Fig. \ref{fig3} we show the maximum contrast between the
intensity of the Fourier peaks, defined as $\Delta _{\max
}\equiv(|c_{1}|^{2}-|c_{0}|^{2})_{\max}$, where the value is
maximized by evaluating the $|c_n|^2$ at the time when $f$ takes
its minimum value (we note that  $f(\tau)$ is given by Eq. (\ref{solu})). We note that the maximum contrast
occurs for $\gamma \approx3.9$; in this case, the zero quasi-momentum
component $c_0(\tau)$ vanishes once during each period
of oscillation $T(\gamma)$.
For values of $\gamma$ greater
than $3.9$ the contrast between the intensities starts to decrease
due to the reduction in tunneling caused by the
nonlinearity-
induced self-trapping.

\subsection{Application to an inhomogeneous condensate}
Here we wish to consider the dynamics for an inhomogeneous condensate, applicable to a condensate initially prepared in a harmonic trap. For the pattern loaded condensate, we use inhomogeneous to refer to the overall spatial envelope of the period three initial condition.
The previous homogeneous theory we have presented is expected to accurately
describe inhomogeneous cases when the initial pattern of population in every third site
extends over many lattice sites i.e. $N_{\rm s}\gg1$ so that mean-field
energy associated with each triplet of sites
$U(n)=\frac{\Lambda}{2}\sum_{i=1}^3|\Psi_{3n+i}|^4$ varies slowly across the system.
Taking as a particular example we choose a gaussian
envelope to the periodic arrangement of atoms into every third
site, so that the initial state is
\begin{eqnarray}
 \Psi_{3n+1}(0)&=&\Psi_{3n+2}(0)=0,\\
 \Psi _{3n}(0)&=&3\left[\frac{2}{\pi N_{\rm s}^{2}}\right]^{1/4}
\exp(-(9n/N_{\rm s})^{2}).
\end{eqnarray}

\begin{figure}[!tbh]
{\centering \includegraphics[width=3.2in]{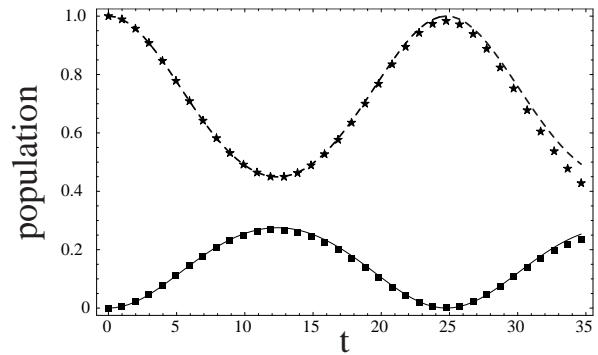}
\par}
\caption{Comparison between the population evolution
 of the central three wells for the
inhomogeneous condensate  and the homogeneous model.
inhomogeneous condensate: stars for the initially
populated well and boxes for the initial empty wells. Homogeneous model: dashed line represents the initially populated wells, and the
solid line represented the initially unpopulated wells. We used $\gamma_{\rm{eff}}$ as the
local mean field energy. The parameters used for the simulation
were $J=0.075E_{\rm{R}}$ and $\gamma_{\rm{eff}}=2.64$. The time is in
computational units: $t\equiv t_{\rm{SI}}\hbar/E_{\rm{R}}$ where $E_{\rm{R}}$
is the recoil energy and $t_{\rm{SI}}$ is time in SI units.
($10\hbar/E_{\rm{R}} \approx 0.7$ms). \label{fig4}}
\end{figure}

For the simulations shown here, the parameters used were
$N=10^{5}$, $N_{\rm s}=76$,  $U = 2.11 \times 10^{-5} E_{\rm R}$
and $J=0.075 E_{\rm R}$, where $E_{\rm R}=h^2/8Ma^2$ is the lattice recoil energy; these correspond to a condensate of $10^5$
atoms of $^{87}$Rb produced in a magnetic trap with axial and
radial frequencies of $9$ and $12$ Hz respectively, and loaded
into a lattice of counter-propagating light $785$nm of depth $4.5E_{\rm{R}}$, with $E_{\rm{R}}=2.2$kHz. These parameters are typical of the
experimental regime, but also lie in a range in which the
homogeneous model is expected to give a fair description.  The
total number of occupied wells and the strength of the on site
interatomic interaction were calculated by preserving the value of
chemical potential of the system upon reduction to one spatial
dimension and by assuming that each of the localized orbitals in
the tight binding description are Gaussian. The hopping rate $J$
was estimated by using Mathieu functions. In Fig. \ref{fig4} we
show the evolution of the population of the central wells
(normalized to one) compared with the homogeneous model with
$\gamma$ taken to be the local mean-field energy
$\gamma_{\rm{eff}}=\Lambda
\sqrt{\frac{2}{\pi}}%
\left( 9/N_{\rm s}\right) $.

Figure \ref{fig4}, shows the results of numerical integration
of the equations of motion and the approximate analytical
solution given by the
quasi-homogeneous model described above.
We see that the numerical and analytical results
agree well at short times, but differ more as
time progresses due to the
different mean-field seen by different wells.

\begin{figure}[!tbh]
{\centering \includegraphics[width=3.2in]{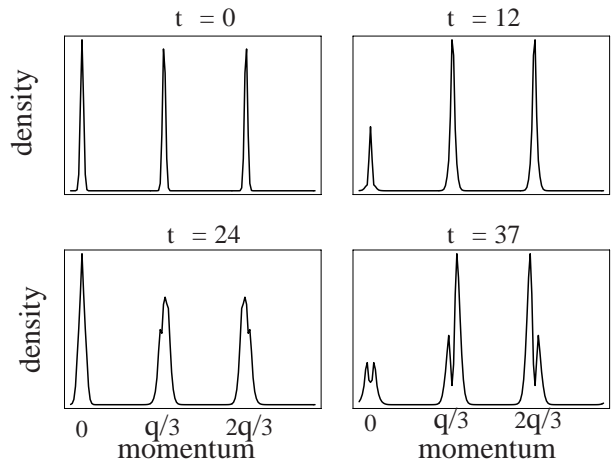}
\par}
\caption{Momentum distribution of the inhomogeneous condensate evaluated
at various times  for the same parameters as used in Fig.\ref{fig4}.  \label{fig5}}
\end{figure}

To understand the disagreement as time evolves, we show in Fig.
\ref{fig5}  the numerical  Fourier spectrum for the inhomogeneous
case evaluated at several different times. The variation of the
intensities of the peaks, which, as shown below, is related to the
spatial tunneling between lattice sites in the presence of the
mean field, can be seen in the plot. Initially all occupied sites
are in phase and the three distinctive momentum peaks have a
narrow width determined by the intrinsic momentum uncertainty of
the condensate envelope.That is the reason why the homogeneous
model fits very well. However, as time progresses the meanfield
variation across the lattice causes the tunneling rates to vary
with position and leads to the momentum peaks broadening. The
homogeneous description then, starts not to be very accurate.

\begin{figure}[!tbh]
{\centering \includegraphics[width=3.2in]{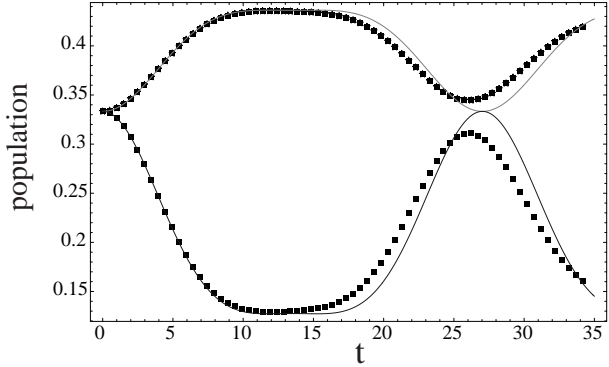}
\par}
\caption{Evolution of momentum peak populations. Upper curves: population
of the $q=\pm2\pi/3$ momentum states. Lower curves: population of the $q=0$
momentum state.
Inhomogeneous condensate (dotted), homogeneous result  (solid line),
where the comparison is made by replacing $\gamma$ by an average
mean field energy $\gamma_{\rm{ave}}\equiv\Lambda
\sum_{n}|\Psi _{n}|^{4}=1.85$. Parameters are the same as in
Fig. \ref{fig4}.  \label{fig6}}
\end{figure}

We note that momentum space signature for
spatially tunneling in the interacting system is still present in
the inhomogeneous case.  This is shown in Fig. \ref{fig6}, where we plot the evolution
of the quantities\ $|c_{3n}(\tau)|^{2}$,
$|c_{3n+1}(\tau)|^{2}$, $|c_{3n+2}(\tau)|^{2}$ $\ $(calculated from
the numerical simulation by partitioning the numerical Fourier
spectrum in three equal non overlapping sections, each centered
around the respective peak and adding the square of the norm of
the Fourier components within each section) {\em vs.} the ones calculated
with the homogeneous model, but using an averaged value
$\gamma_{\rm{ave}}\equiv\Lambda \sum_{n}|\Psi _{n}|^{4}$ instead
of  $\gamma=\Lambda \rho$. It can be observed
that the predictions of the simple model are in very good
agreement with the numerical results when the three peaks of the
spectrum are well defined. For longer times, the width of the
Fourier peaks increases, until a point when they split. At this
point the quantities $|c_{3n}(\tau)|^{2}$, $|c_{3n+1}(\tau)|^{2}$,
$|c_{3n+2}(\tau)|^{2}$ are not meaningful anymore. Because the
parameters used for the numerical calculations were chosen to be
experimentally achievable, and as shown in the plots the model
predictions are fair at least for one period of oscillation, we
conclude that Fourier distribution can be used as a signature of
the mean-field quantum tunneling inhibition.

\section{Dynamics with a constant external force}
\label{secextpot}
\subsection{Homogeneous three state model}
In this section we consider the dynamics of a periodically loaded
condensate in the presence of  a linear external potential,
corresponding to a uniform force parallel to the lattice. In what
follows we assume that the force is sufficiently weak that band
excitations due to Landau-Zener tunneling is negligible, so that a
tight binding picture of the lowest band is sufficient to describe
the dynamics.
 In this case the evolution equation differs from what we considered in the
previous section
 by the term $E_{n}$ in Eq.
(\ref{tres}) taking the form $E_{n}=n\xi$, where $\xi$ is the
potential difference between lattice sites (in units of the
hopping matrix element $J$). Taking the initial conditions
(\ref{IC1})-(\ref{IC2}), and transforming the Wannier amplitudes
as $\Psi_{3n+j}(\tau)=\tilde{\Psi}_{3n+j}(\tau)e^{-i3n\xi \tau}$
$(j=0,1,2)$, we obtain the evolution equations
 \begin{eqnarray}
i\frac{\partial \tilde{\Psi}_{3n}}{\partial \tau} &=&-(\tilde{\Psi}%
_{3n-1}e^{i3\xi \tau}+\tilde{\Psi}_{3n+1})\nonumber\\
&&+\Lambda |\tilde{\Psi}_{3n}|^{2}\tilde{\Psi}_{3n}, \\
i\frac{\partial \tilde{\Psi}_{3n+1}}{\partial \tau}
&=&-(\tilde{\Psi}_{3n+2}+%
\tilde{\Psi}_{3n})+\xi \tilde{\Psi}_{3n+1}\nonumber\\
&&+\Lambda |\tilde{\Psi}_{3n+1}|^{2}\tilde{\Psi}_{3n+1}, \\
i\frac{\partial \tilde{\Psi}_{3n+2}}{\partial \tau}
&=&-(\tilde{\Psi}_{3n+1}+%
\tilde{\Psi}_{3n+3}e^{-i3\xi \tau})+2\xi \tilde{\Psi}_{3n+1}\nonumber\\
&&+\Lambda |\tilde{\Psi}_{3n+1}|^{2}\tilde{\Psi}_{3n+1},
\end{eqnarray}
Assuming periodic boundary conditions, the periodicity of the
initial conditions and equations of evolution allow considerable
simplification from the full set of $3T$ coupled equations. In
particular these assumptions mean that every third Wannier
amplitude evolves identically (i.e.
$\tilde{\Psi}_n=\tilde{\Psi}_{n+3}$) and so the evolution of the
system can hence be reduced to the three independent equations
\begin{eqnarray}
i\frac{\partial \tilde{\Psi} _{0}}{\partial \tau} &=&-(\tilde{\Psi} _{2}e^{i3\xi
\tau}+\tilde{\Psi}
_{1})+\Lambda |\tilde{\Psi} _{0}|^{2}\tilde{\Psi} _{0},\label{Feq1} \\
i\frac{\partial \tilde{\Psi} _{1}}{\partial \tau} &=&-(\tilde{\Psi} _{2}+\tilde{\Psi}
_{0})+\xi
\tilde{\Psi} _{1}+\Lambda |\tilde{\Psi} _{1}|^{2}\tilde{\Psi} _{1},\label{Feq2} \\
i\frac{\partial \tilde{\Psi} _{2}}{\partial \tau} &=&-(\tilde{\Psi} _{1}+\tilde{\Psi}
_{0}e^{-i3\xi \tau})+2\xi \tilde{\Psi} _{2}\nonumber \\
&&+\Lambda |\tilde{\Psi} _{2}|^{2}\tilde{\Psi} _{2}\label{Feq3},
\end{eqnarray}
where the new amplitudes map onto the original set according to
$\tilde{\Psi}_0\leftrightarrow \{\tilde{\Psi}_{3n}\}$,
$\tilde{\Psi}_1\leftrightarrow\{\tilde{\Psi}_{3n+1}\}$,
and $\tilde{\Psi}_2\leftrightarrow\{\tilde{\Psi}_{3n+2}\}$, and
obey the normalization condition
\begin{equation}
\sum_{j=0}^2|\tilde{\Psi}_j|^2=\frac{1}{T}.
\end{equation}
The equations of motion (\ref{Feq1})-(\ref{Feq3}) are more
difficult to treat analytically than the case considered in the
last section due to the presence of a linear potential. In this
paper we derive an analytic solution for a noninteracting
condensate (i.e. $\Lambda $=0), which provides valuable insight
into the complicated tunneling dynamics the system exhibits in the
absence of nonlinearity, yet should furnish a good description for
dilute condensates satisfying $\gamma\ll1$. For the nonlinear
regime we present numerical results to illustrate typical
behavior.

\subsection{Analytic solution for linear dynamics}

Defining the vector ${\bf x}(t)=(\tilde{\Psi} _{0}(\tau),\tilde{\Psi}
_{1}(\tau),\tilde{\Psi}_{2}(\tau))$, and using the transformation $\ {\bf
y}(\tau)=P(\tau){\bf x}(\tau)$, where $P(\tau)$ is the unitary
matrix
\begin{equation}
P(\tau)=\left(
\begin{array}{ccc}
1 & 1 & 1 \\
e^{-i\xi \tau} & e^{-i(\xi \tau-\frac{2\pi }{3})} & e^{-i(\xi\tau+\frac{%
2\pi }{3})} \\ e^{-i2\xi\tau} & e^{-i(2\xi\tau+\frac{2\pi }{3})} &
e^{-i(2\xi\tau-\frac{2\pi }{3})}
\end{array}\right),
\end{equation}
the linear version of Eqs. (\ref{Feq1})-(\ref{Feq3}) can be
decoupled, directly yielding the solutions

\begin{eqnarray}
\tilde{\Psi} _{n}(\tau) &=&\frac{e^{-in\xi \tau}}{3\sqrt{T}}\Big(
e^{-i\Delta _{0}(\tau)}+e^{i\left( \frac{2n\pi }{3}-\Delta
_{1}(\tau)\right)}  \nonumber \\
& & +e^{-i\left( \frac{2n\pi }{3} +\Delta
_{2}(\tau)\right) }\Big), \label{po1}
\end{eqnarray}

\noindent where we have defined the phase terms
\begin{eqnarray}
\Delta _{n}(\tau) &=&-\frac{2}{\xi }\left[ \sin (\xi \tau -\frac{2n\pi
}{3})+\sin (%
\frac{2n\pi }{3})\right]   \label{ev1}
\end{eqnarray}

\noindent for $n=0,1,2$.

These solutions for the spatial amplitudes $\tilde{\Psi} _{i}(\tau)$ can
be most easily understood by considering a Bloch state
decomposition of the condensate wavefunction. The nature of our
system allows us to construct an analytic form for the initial
wavefunction. Because the system has a three lattice site period
and is assumed to be in the lowest band, the wavefunction can be0
expressed as a superposition of three Bloch waves (of the lowest
band) which are symmetrically spaced in quasimomentum. Assuming
the condensate initially has a total crystal momentum of zero, at
this time the wavefunction must be of the form
\begin{equation}
\Phi (x,0)=\alpha_0 \varphi _{0}(x)+\alpha_+  \varphi _{q/3}(x)+\alpha_-
\varphi _{-q/3}(x)
\end{equation}
where $\varphi _{k}(x)$ is a Bloch state  with quasimomentum $k$,
and the $\alpha$ are complex constants
determined by the lattice depth, with $| \alpha_+|=| \alpha_-|$.

The action of an external force on a Bloch state causes it to
linearly change its quasimomentum in time according to
\begin{equation}
k(\tau)=-\frac{\xi }{a}\tau+k(0).
\end{equation}
The periodicity of the Bloch dispersion relation in $k$, and hence
of the  group velocity of the Bloch wave, gives rise to the
well-known phenomenon of Bloch oscillations \cite{mermin1976}. For
the case we are considering here, the system consists of three
Bloch states whose quasimomenta will translate in unison under the
action of the external force. During this evolution each state
accumulates phase at a rate determined by the instantaneous Bloch
energy, i.e.
\begin{equation}
\Delta _{n}(\tau) = \int_{0}^{\tau}E(k_n(s))ds,
\end{equation}
where $k_n(\tau)=-\xi\tau/a+k_n(0)$ is the quasimomentum of Bloch
state $n$ at time $t$. In the tight binding approximation the
dispersion relation for the Bloch states has the analytic form
\begin{equation}
E(k)=-2 \cos(ka),\label{dispersionTB}
\end{equation}
for which $\Delta_n(\tau)$ can be evaluated,and  yields the
results given in Eqs. (\ref{ev1})).  The wavefunction evolution in
the Bloch basis is
\begin{eqnarray} \Phi (x,\tau)&=&\alpha_0 \varphi _{-\frac{\xi
}{a}t}(x)e^{-i\Delta _{0}(\tau)}+\alpha_+ \varphi _{-\frac{\xi
}{a}t+\frac{q}{3}}(x)e^{-i\Delta _{1}(\tau)}\nonumber\\&&+\alpha_-
\varphi _{-\frac{\xi }{a}t-\frac{q}{3}}(x)e^{-i\Delta _{2}(\tau)}.
\end{eqnarray}

\noindent From this solution we can obtain solutions for the evolution of
the spatial amplitudes Eqs. (\ref{ev1}). To do this we expand the
Bloch states in terms of Wannier functions according to $\varphi
_{nk}(x)=\sum_{n}e^{ikna}\phi _{n}(x)$ and make use of Eq.
(\ref{dos}). Note: we take all $\alpha = \frac{\rho}{3}$
as determined by the initial conditions, Eqs.(\ref{IC1}) and
(\ref{IC2}).

\subsubsection{Bloch Oscillations}
The evolution of the spatial amplitudes, and in particular the
population in each well is then determined by the interference of
the Bloch phases $\Delta_n$. These functions are all periodic in
time with period $\tau_B=2\pi/\xi $ (in units of $\tau =tJ/\hbar)$. This
is the normal Bloch oscillation period, and gives the time scale
over which the quasimomenta of the Bloch states develop by exactly
a lattice vector.

\subsubsection{Small $\xi$ solution - Non classical transport}
To understand the dynamics, we first start by considering the case
when $\xi $ is small. For this case, the population in the wells
is given by
\begin{eqnarray}
|\tilde{\Psi} _{0}(\tau)|^{2} &=&\frac{1}{3 N_{\rm s}}\Big( 5+4\cos (3\tau
)+ \xi^{2} h(\tau) \Big) \\
|\tilde{\Psi} _{1}(\tau)|^{2} &=&\frac{1}{3 N_{\rm s}}\Big( (2 + 3\xi
)[1-\cos(3\tau)- \frac{\xi^{2}}{2} h(\tau) ]\Big) \\
|\tilde{\Psi} _{2}(\tau)|^{2} &=&\frac{1}{3 N_{\rm s}}\Big( (2 - 3\xi
)[1-\cos(3\tau)- \frac{\xi^{2}}{2} h(\tau)]\Big)
\end{eqnarray}
where
\begin{equation}
h(\tau)= \frac{\tau^{3} }{2}\left(6\tau +3\tau \cos(3\tau) -4\sin (3\tau )\right).
\end{equation}
The above solution shows that when the force is applied, the
degeneracy in the population of the wells represented by $\tilde{\Psi}_1$
and $\tilde{\Psi}_2$ is lifted. For $\xi>0$, atoms in $\tilde{\Psi}_0$  start to
tunnel to $\tilde{\Psi}_1$ more rapidly than to $\tilde{\Psi}_2$. This should be
compared with the results in the absence of the force, where
$\tilde{\Psi}_1$ and $\tilde{\Psi}_2$ behave identically.
 Thus, in this weak
limit, the effect of a linear potential to
enhance the tunneling from the initial populated $3n$ wells to
their $3n+1$ neighbors
ones, making the system closer to resonance,
in the sense of Eq. (\ref{nl2state}), where the
resonance condition results in the
initially
populated wells becoming empty at some later time.
It is interesting to note that the
system exhibits "nonclassical'' dynamics whereby the atoms start
to tunnel in the direction opposite the direction of the force
(this statement applies even when the external
 field is not weak).
\begin{figure}[!tbh]
{\centering \includegraphics[width=3.2in]{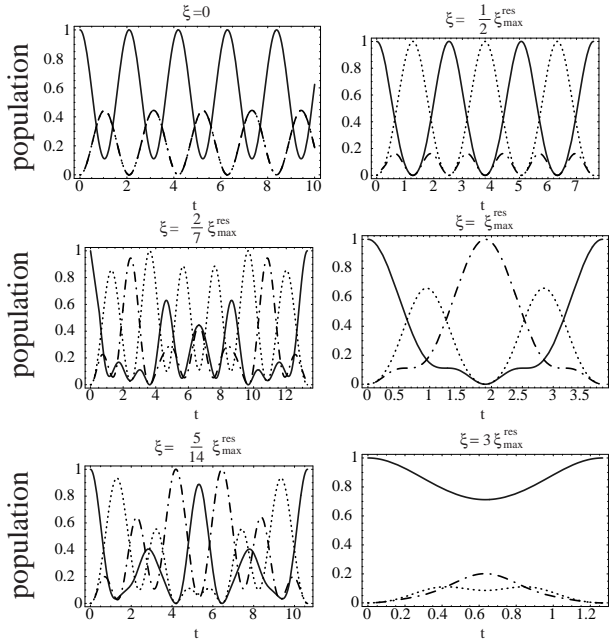}
\par}
\caption{Evolution of the normalized population for different values of $\xi$.
One Bloch period is shown in the plots except $\xi=0$ where the period is infinite.
Solid line: $|\tilde{\Psi}_{3n}|^2$, dotted line: $|\tilde{\Psi}_{3n+1}|^2$, dashed line: $|\tilde{\Psi}_{3n+2}|^2$.  The
"nonclassical" motion can be seen where the $3n\!+\!1$-well populations increase more rapidly that the populations of the $3n\!+\!2$-wells.
   It can also be seen in the plots that $\xi=\xi_{\rm{max}}/2$ and
$\xi=\xi_{\rm{max}}$ are resonant values.\label{fig7}}
\end{figure}
\begin{figure}[!tbh]
{\centering \includegraphics[width=3.2in]{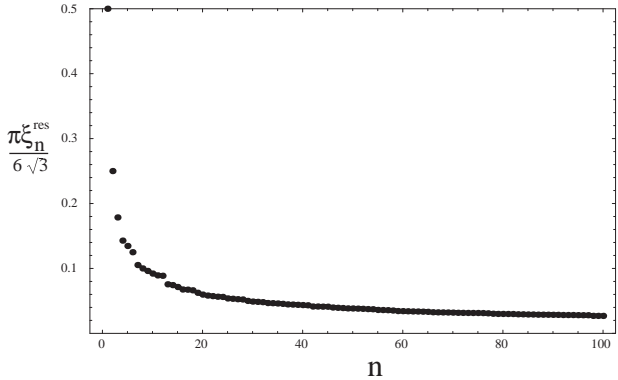}
\par}
\caption{The spectrum of values of external force, ordered in decreasing magnitude, for which a population resonance occurs i.e. the values of  $\xi$ for which
$\tilde{\Psi}_0$  periodically disappears  \label{fig8}}
\end{figure}
\subsubsection{Resonances}
In Fig. \ref{fig7} we show the temporal evolution of the spatial amplitudes
$\tilde{\Psi}_i$ for a range of values of $\xi$. For certain choices of
$\xi$ the initially occupied $\tilde{\Psi}_0$ amplitude periodically
disappears  - we refer to these as resonances. By requiring
$\tilde{\Psi}_0=0$ in Eq. (\ref{po1}) we obtain the following conditions
on the phases for these resonances
\begin{eqnarray}
\Delta _{1}(\tau)-\Delta _{0}(\tau) &=&\pm \left(\frac{\pi}{3}+2\pi n\right)
\label{rescond1}\\
\Delta _{2}(\tau)-\Delta _{1}(\tau) &=&\pm
\left(\frac{\pi}{3}+2\pi m\right)\label{rescond2}
\end{eqnarray}
where $n$ and $m$ are integers and the same sign choice on the
right hand side must be made for both equations. When these
conditions are satisfied the population is not shared between the
adjacent sites, but preferentially tunnels to one of the
neighbors. In particular the $+$sign choice in Eqs.
(\ref{rescond1}) and (\ref{rescond2}), gives the phase condition
for complete tunneling into $\tilde{\Psi}_1$. Similarly the $-$sign case
gives the condition for complete tunneling into $\tilde{\Psi}_2$. Solving
for the values of $\xi$ for which  Eqs. (\ref{rescond1}) and
(\ref{rescond2}) hold, gives the spectrum
\begin{equation}
\xi _{nm}^{\rm{res}}=\pm \frac{6\sqrt{3}}{\pi }\frac{(1+3n)}{%
(1+3n)^{2}+3(n+2m+1)^{2}}.
\end{equation}
which is shown in Fig. \ref{fig8}. For large $|n|$ or $|m|$ the resonant
values of $\xi$ are close together, and become spaced further
apart as the magnitude of $n$ decreases.

In the absence of an applied force (i.e. $\xi =0$) the phase terms
are time-independent with $\Delta _{0}=-2$, $\Delta _{1}=-1$ and
$\Delta _{2}=-1$ so that the resonance conditions can never be
achieved. When the external force is applied the phases oscillate
at a rate that increases with $\xi$ and an amplitude that
decreases with $\xi$. The largest value of $\xi$ for which a
resonance can be found is $\xi _{\max
}^{\rm{res}}\equiv 6\sqrt{3}/2\pi$, since for values of $\xi $
greater than this the amplitudes of the phase oscillations
$\Delta_i$ are so small  that they cannot satisfy the condition
for population resonance. In the regime $\xi>\xi_{\max}^{\rm{res}}$,
$|\tilde{\Psi}_0|$ exhibits only one (nonzero) minima per Bloch period and
the dynamics of the system is dominated by the Bloch oscillations.

\subsubsection{Large Force Limit}
For $\xi> \xi_{\max}^{\rm{res}}$ the system exhibits a population
imbalance as the Bloch oscillation suppresses the ability of the
system to tunnel between sites. In the limit $\xi\gg\xi _{\max
}^{\rm{res}}$ the population in the wells is described by rapid, small
amplitude oscillations around its initial value,
\begin{eqnarray}
|\tilde{\Psi} _{0}(\tau)|^{2} &\longrightarrow &\frac{3}{N_{\rm s}}\left( 1-\frac{8}{\xi
^{2}}%
\sin ^{2}\left( \frac{\xi \tau }{2}\right) \right)  \\ |\tilde{\Psi}
_{1}(\tau)|^{2} &\longrightarrow &\frac{12}{N_{\rm s}}\frac{1}{\xi
^{2}}\sin ^{2}\left( \frac{\xi \tau }{2}\right)  \\ |\tilde{\Psi}
_{2}(\tau)|^{2} &\longrightarrow &\frac{12}{N_{\rm s}}\frac{1}{\xi
^{2}}\sin ^{2}\left( \frac{\xi \tau }{2}\right)
\end{eqnarray}

\noindent The case of $\xi = 3 \xi^{\rm{res}}_{\max}$ depicted in Fig. 8
shows an approach to this behavior.

\subsection{Numerical results for the non-linear case}
\begin{figure}[!tbh]
{\centering \includegraphics[width=3.2in]{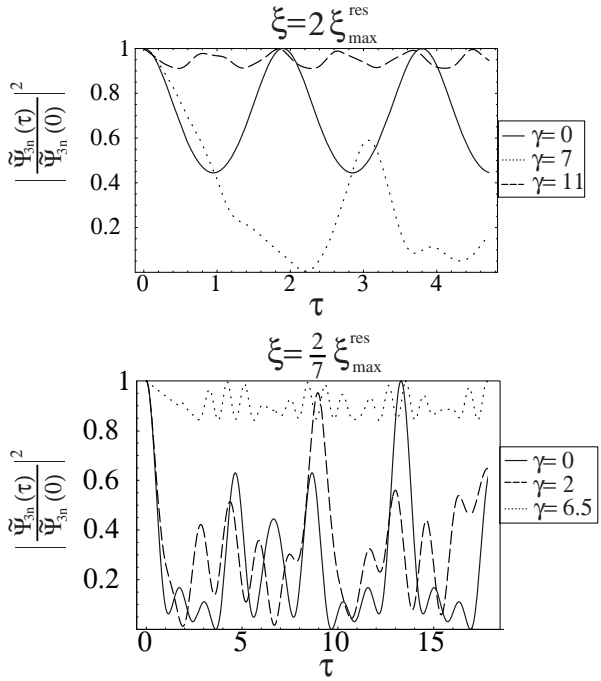}
\par}
\caption{Effects of interactions on generalized Bloch oscillations for the pattern loaded system. Evolution of $|\tilde{\Psi}_n|^2$ for various interaction strengths. Upper plot: $\xi=2\xi^{\rm{res}}_{\max}$.
Lower plot: $\xi=2\xi^{\rm{res}}_{\max}/7$.
\label{fig9}}
\end{figure}
Because of the difficulty of solving analytically the equations of
motion the nonlinear equations were solved numerically.
Although the
dynamics in this situation is
considerably more complicated
than in the linear case, the resonance picture
still
gives us useful guidance concerning the
expected behavior.
 In general,
there exist critical values of $\xi $ and $\Lambda$
above which no resonances occur.  Furthermore, nonlinearity
tends to destroy the periodicity of the Bloch oscillations
that is present in the noninteracting case.  For example,
in Fig. \ref{fig9}, we see that for $\xi = 2 \xi^{\rm{res}}_{\max}$, the
introduction of interactions brings the system into resonance,
but that for large interaction strength, nonresonant
behavior is restored.  For $\xi = 2 \xi^{\rm{res}}_{\max}/7$, on the other
hand, the introduction of interactions eventually draws the
system out of resonance.

Concerning the momentum distribution, similarly to the untilted case,
interatomic interactions induce time variation of the momentum
intensities. The contrast between momentum
components vanishes at $\Lambda =0$.  As $\Lambda$ increases,
the contrast increases to a maximum value, then
eventually decreases towards zero when macroscopic imbalance is
achieved, analogous to what is seen in Fig. \ref{fig3} for the
untilted case.  Because the external field
together with the nonlinearity breaks down the periodicity
of time evolution, the
dynamics of the momentum distribution is quite complex.

\begin{figure}[!tbh]
{\centering \includegraphics[width=3.2in]{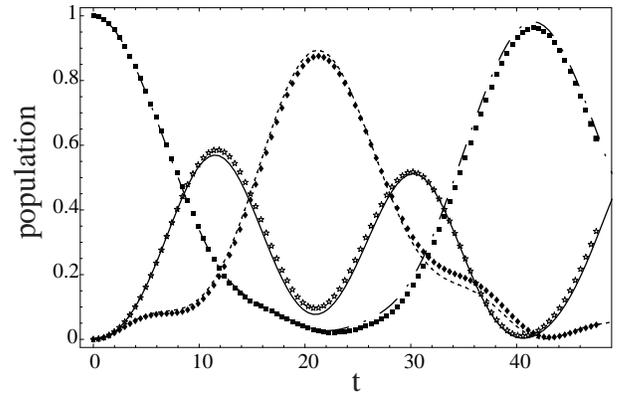}
\par}
\caption{ Comparison between the evolution of a inhomogeneous condensate
with the homogeneous result. Inhomogeneous condensate: $|\tilde{\Psi}_{0}|^2$ (boxes),
  $|\tilde{\Psi}_{1}|^2$ (stars), and
 $|\tilde{\Psi}_{2 }|^2$ (diamonds). Homogeneous case: $|\tilde{\Psi}_{3n}|^2$ (dash-dot line),
  $|\tilde{\Psi}_{3n+1}|^2$ (solid line), and
 $|\tilde{\Psi}_{3n+2 }|^2$ (dotted line), where we have taken $\gamma$ as the local mean field energy.
 The parameters used were $J=0.075E_{\rm{R}}$, $\xi=2$ and $\gamma_{\rm{eff}}=1.59$
\label{fig10}}
\end{figure}

\begin{figure}[!tbh]
{\centering \includegraphics[width=3.2in]{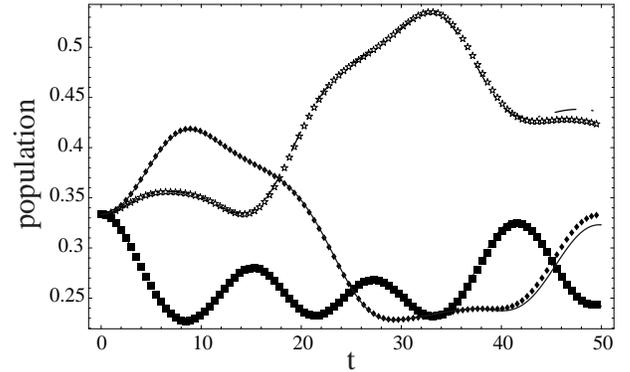}
\par}
\caption{ Evolution of the momentum peak populations.
$q=0$ (dashed line, squares), $q=-2\pi/3$ (dash-dot line, stars), $q=2\pi/3$
(solid line, diamonds).
Inhomogeneous condensate (dotted curves),
homogeneous model (lines) using the same parameters as those
 in Fig. \ref{fig10}, but replacing $\gamma$ by an average value
$\gamma_{\rm{ave}}=1.11$ for the homogeneous model. \label{fig11}}
\end{figure}

To compare
the predictions of the homogeneous model in the presence of a
linear external potential to a more realistic case, we again
solved numerically the discrete nonlinear Schr\"odinger equation
for a condensate loaded every three
lattice sites but, instead of being homogeneous, initially with a
Gaussian profile. Very good agreement between the model and the
numerical results was found for short times and modest
mean-field energies, if, as in the untilted case, we use an effective mean
field energy. In Figs. \ref{fig10} and \ref{fig11} we present a comparison between the
evolution of the normalized population at the central wells
found numerically and the prediction of the model for the parameters:
  $NU=4.8 E_{\rm R}$, $N_{\rm s}=290$, $J=0.075 E_{\rm R}$
and $\xi =2$ with computational time of $50$ is
about $3.5$ ms. For longer times, the model predictions start to
disagree with numerical simulation due to the spread of the three
Fourier peaks. We observe that the dynamics are much more sensitive to inhomogeneous
effects in the presence of an external force, and for the more inhomogeneous
initial state used in Figs. 4-6
(which extends over $N_{\rm{s}}=76$ sites), the inhomogeneous result much more rapidly
departs from the homogeneous prediction than the example presented here.


\section{Conclusion}

We have studied the dynamics of a condensate loaded in an optical
lattice in such a way that initially only every third well in the
lattice was occupied. We considered two cases, one with only the
optical potential, and one with a superimposed constant external
field. It was found that in both cases interatomic interactions
can cause macroscopic quantum self
trapping, which is a self maintained population imbalance across
the wells. The analysis was based on a tight binding model
that assumes that at $\tau=0$ all the $3n$ wells were identically
populated. Using this model, we studied the different regimes
where the self trapping phenomenon occurs and estimated the
critical parameters for each case. We have developed an analytic solution
for the dynamics of a periodic initial condition in the absence of an external
potential.
For this case the symmetry of the system reduces the discrete nonlinear
Schr\"odinger equation to a two-mode problem with an elliptic function solution.
We have verified the usefulness of the analytic solution by comparing it to
numerical solutions of the discrete nonlinear Schr\"odinger equation for more
general initial conditions.

We have demonstrated that meanfield effects
cause the momentum distribution to vary with time, and have shown how
this variation relates to the spatial tunneling in the system.
This result suggests that the temporal variation in the
time-of-flight density distribution (which approximately corresponds to the
\emph{in-situ} momentum distribution) would be a useful
experimental signature of the spatial tunneling.

In the presence of an
external potential and neglecting interactions, it was found that
the spatial population amplitudes evolve periodically in time, with the
oscillation period determined by the force parameter
$\xi$. We have shown that these dynamics can be understood as the
interference of three quasimomentum states Bloch oscillating in unison.
When interactions
are taken into account, the periodicity of the system is
destroyed and the atoms exhibit more complicated dynamics.

\section*{Acknowledgements}
This work was supported in part by the Advanced Research and Development
Activity and by the U.S. National Science Foundation under
grant PHY-0100767.

\end{document}